\documentclass[namedreferences,hyperref,optionalrh,solaromanenum]{spr-sola}

\usepackage{graphicx}                    
\usepackage{color}                       

\chardef\us=`\_

\usepackage{lscape}

\definecolor{mygreen}{RGB}{0, 185, 118}

\begin{document}

\begin{frontmatter}

\title{A Multi-viewpoint CME Catalog Based on SoloHI Observed Events}

%
\author[addressref={1,2}, email={cecilia.maccormack@nasa.gov}]{\snm{C. Mac Cormack}}
\author[addressref={3}]{\inits{3}\fnm{S. B. Shaik}}
\author[addressref={4}]{\inits{4}\fnm{P. Hess}}
\author[addressref={4}]{\inits{4}\fnm{R. Colaninno}}
\author[addressref={1}]{\inits{1}\fnm{T. Nieves-Chinchilla}}

%
\runningauthor{C. Mac Cormack et al.}
\runningtitle{A Multi-viewpoint CME catalog based on SoloHI observed events}

   \address[id={1}]{Heliospheric Physics Laboratory, Heliophysics Science Division, NASA Goddard Space Flight Center, 8800 Greenbelt Rd., Greenbelt, MD 20770, USA}

   \address[id={2}]{The Catholic University of America, Washington, DC 20064, USA}

   \address[id={3}]{George Mason University, Fairfax, VA 22006, USA.}

   \address[id={4}]{U.S. Naval Research Laboratory, Washington, D.C. 20375, USA}


\begin{abstract}

Coronal Mass Ejections (CMEs) are significant drivers of geomagnetic activity, and understanding these structures is critical to developing and improving forecasting tools for space weather. The Solar Orbiter (SolO) mission, with its comprehensive set of remote sensing and in-situ instruments, along with its unique orbit, is significantly advancing the study of the CMEs and other structures in the heliosphere. A critical contribution to the study of CMEs by SolO is the observations from the Solar Orbiter Heliospheric Imager (SoloHI). SoloHI observes photospheric visible light scattered by electrons in the solar wind and provides high-resolution observations of the corona and heliosphere. The resolution and vantage point offered by SoloHI make it uniquely well-suited to study CME evolution in the heliosphere. To contribute to the science goals of SolO, we present the initial release of a living CME catalog based on SoloHI observations during its initial years of observations, with a multi-viewpoint focus. We catalog 140 events detected by SoloHI during the period of January 2022 until April 2023. For each event detected by SoloHI, we present available in-situ data and remote sensing observations detected by other missions. With the available observations, we identify the source region of the CME and describe its main characteristics, track the CME through the coronagraphs and heliospheric imagers, and provide in-situ detection when possible. We also provide a morphological classification and observations quality parameter based on the SoloHI observations. Additionally, we cross-check with other available CME catalogs and link to the event description provided by the Space Weather Database Of Notifications, Knowledge, Information (DONKI) catalog developed at the Community Coordinated Modeling Center (CCMC). In this article, we describe the features of the SoloHI CME catalog and the methods used to generate the entries. We also present a statistical study of the morphological classification of the cataloged CMEs in the SoloHI observations, building up on the previous studies that classify the events observed by LASCO coronagraphs. We provide various observing scenarios with SoloHI observations to demonstrate the contribution that this catalog offers to the scientific community to explore the new observing viewpoint of CMEs with the SolO mission.
\end{abstract}

%

\end{frontmatter}

\section{Introduction}

Coronal Mass Ejections (CMEs) are large-scale structures that can be observed by both remote sensing (RS) and in-situ (IS) instrumentation. CMEs are responsible for heliospheric variability and are among the main drivers of space-weather activity. As they evolve and propagate, CMEs change the conditions in the interplanetary medium by perturbing it and filling it with energetic particles and enhanced magnetic fields. Understanding these processes is important for predicting events that could cause potential damage to technological systems, space missions, and astronauts.

Multi-viewpoint studies are the key to understanding the global evolution of CME events and tracking them throughout the heliosphere is a task fully dependent on space mission coverage. Our understanding of CMEs' heliospheric variability has been improved significantly by studies that incorporate multiple viewpoints data from {different heliospheric mission} \citep[e.g.][]{braga_2022,niembro_2023,davies_2024}. 
As the number of these space-based missions is increasing, the effort in the community is building toward the integration of all the data for a comprehensive understanding of the observed phenomena. {The latest contributions to the range of spacecraft monitoring corona and heliospheric activity are} the new generation of solar and planetary missions, e.g., Parker Solar Probe \citep[PSP:][]{fox_2016}, BepiColombo \citep[Bepi:][]{benkhoff_2021}, and Solar Orbiter \citep[SolO:][]{muller_2020}. These missions are in novel orbits with the closest approach to the Sun of well within 1 au and enhanced resolution.

{While previous generations of synoptic missions provide regular data, enabling the study of long-term statistical trends across wide fields of view (FOVs), the new generation of solar missions such as PSP, SolO, are encounter missions, yielding high-cadence, high-resolution data during specific orbital periods. Although these data are groundbreaking for global multi-viewpoint analysis, they are less suited to general statistics and descriptions of CME properties. Combining synoptic and encounter mission data unlocks their full potential and it is crucial in this new mission era for fostering a comprehensive understanding of CME evolution.}


{Moreover,} with the combination of remote sensing (RS) and in-situ instruments on the same spacecraft, the SolO mission incorporates the multi-viewpoint perspective in its science mission objectives \citep{muller_2020}. Among the six remote sensing instruments on board the SolO mission, the Solar Orbiter Heliospheric Imager \citep[SoloHI:][]{howard_2020} allows us to study the inner heliosphere by observing white light scattered by electrons in the solar wind. SoloHI observes a FOV of 40$^{\circ}$ of elongation starting at 5$^{\circ}$ off the east of the Sun. It consists of four tiles combined together in a pinwheel fashion to form one whole detector. During perihelion, SoloHI observes a range of heights up to 60$\,R_{\odot}$, {reaching a FOV larger than the C3 telescope of the coronograph Large Angle and Spectrometric Coronagraph Experiment \citep[LASCO]{brueckner_1995} on board the Solar and Heliospheric Observatory \citep[SOHO]{domingo_1995} mission, and a better effective resolution than the heliospheric imager HI-1 part of the Sun-Earth Connection Coronal and Heliospheric Investigation \citep[SECCHI]{howard_2008} suite on board the Solar TErrestrial RElations Observatory \citep[STEREO]{kaiser_2008}.}

{On the other hand, the Wide-field Imager for Parker Solar Probe \citep[WISPR:][]{vourlidas_2016} is the only remote-sensing instrument onboard PSP. Unlike SoloHI, WISPR observes at the west of the Sun with respect to PSP, covering the range of elongation angles from 13.5$^{\circ}$ to 108$^{\circ}$, and provides data when the spacecraft is located in a heliocentric distance lower than 0.25$\,$au. As SoloHI and WISPR instruments present several similarities with high-resolution white-light observations, a joint effort can unveil information about the internal structures of CMEs never seen before.}


{{Moreover,} while STEREO-HI enabled unprecedented CME measurements at higher heights and provided regular data for robust statistical analysis of CME parameters, SoloHI offers exceptional resolution for monitoring these heights during certain periods.} \citet{hess_2023} performed a comparative study of a sequence of CMEs during the first SoloHI science perihelion and highlighted improvements and better-resolved structures that the instrument can detect compared to previous generations of imagers. With the higher-resolution white-light images, SoloHI allows us to track specific features detected before the CME erupts through the solar corona and the heliosphere. Moreover, SoloHI will take advantage of the out-of-ecliptic orbit planned for the SolO mission and will provide an additional POV (point of view) in the multi-view analysis of the early evolution of the CMEs at different latitudes.
 
While all instruments onboard SolO are designed to have a significant impact on the study of the Sun and heliospheric activity, SoloHI is a particularly powerful tool to observe and understand CME evolution from the corona into the heliosphere. The other RS instruments are mainly focused on the solar disk and low corona and can provide valuable insights into CME onset and initiation, among many other important science objectives. SoloHI is the only instrument onboard SolO that regularly observes beyond {$5.25\,$au, imaging the overlap between what is generally considered the transition between coronal and heliospheric heights. Then, SoloHI can help us better understand the evolution of CMEs from an organized structure near the Sun into a large interplanetary structure interacting with the solar wind in the heliosphere.}



Since SoloHI, under nominal observing conditions, points to the east of the Sun, it is unlikely to obtain measurements from the in-situ instruments on board SolO for the CMEs that appear most clearly in the SoloHI FOV. However, SoloHI provides invaluable information to describe the events in collaboration with all the other {heliopheric and planetary missions.} 

To maximize the return of the SoloHI instrument and coordinate with the other missions, a thorough accounting of the events that have been seen in the images is needed. In this work, we present a catalog based on the events detected by SoloHI. This living catalog will be constantly updated with the new observations provided by the instrument. {The ability to study a particular CME's velocity, position angle, and angular width in SoloHI data is complicated by its variable FOV and effective resolution. SoloHI makes significant contributions by providing higher spatial and temporal resolution when combined with other synoptic heliospheric observations for context, enhancing our overall understanding.} Aligned with the collaborative philosophy of the SolO mission, this catalog will be an inclusive catalog that integrates observations provided by other missions to globally characterize CMEs and collect as much relevant data in one place as possible.

{Integrating SoloHI observations to the data provided by the suite of spacecraft monitoring the heliosphere is science-enabling and represents a significant tool for global heliospheric analysis. The comprehensive SoloHI CME catalog offers a powerful resource for heliospheric research, facilitating global perspectives and multi-viewpoint analysis for the scientific community.}

In Section \ref{cats}, we present a brief overview of existing, relevant catalogs. In Section \ref{SoloHIdesc}, we describe different scenarios of spacecraft configuration where SoloHI plays an important role by providing observations to the community from an important and unique location. In Section \ref{SoloHIcat}, we describe the content of the catalog and all the sources that were used during its population. In Section \ref{Morpho}, we present a general morphological classification of the events detected from January 2022 through April 2023. In Section \ref{CasesOfStudy}, we perform a detailed analysis of three selected events where SoloHI plays a crucial role in the observing configuration, using the information provided by the catalog. Finally, in Section \ref{Conc}, we summarize the results and conclusions.

\section{Overview of existing catalogs}
\label{cats}

Many CME studies tend to be focused on either in-situ or remote sensing observations. While remote sensing observations allow us to describe the origin and the initial stages of the evolution of CMEs, in-situ measurements provide a unique series of samples of the magnetic field, plasma properties, energetic particles, and internal matter composition of these structures. Both kinds of measurements together allow us to understand the CME's evolutionary processes, i.e., rotation, distortion, deflection, and expansion, among others (see \citet{manchester_2017}). By combining the large amount of data provided by every spacecraft in {orbit}, many CMEs can be tracked from their origin until 1$\,$au. 

Although many catalogs are available online that help us to describe the observations detected by particular instruments, most of them do not provide the complementary information needed to understand the evolutionary processes of any particular CME. 

One of the most comprehensive real-time catalogs available to the community is the Space Weather Database Of Notifications, Knowledge, Information (\href{https://ccmc.gsfc.nasa.gov/tools/DONKI/}{DONKI}\footnote{https://ccmc.gsfc.nasa.gov/tools/DONKI/}), developed at the Community Coordinated Modeling Center (\href{https://ccmc.gsfc.nasa.gov}{CCMC}, \footnote{https://ccmc.gsfc.nasa.gov}) and populated by daily forecasting of the NASA Moon to Mars (M2M) Space Weather Analysis group. The M2M team collects daily information on Sun's activity and its impact on the {different heliospheric and planetary missions}. In this catalog, the community can find a list of events within a selected period, source region information, modeled CME evolution parameters, and the predicted impact on different spacecraft or planets. However, the catalog does not corroborate the arrival at the spacecraft or link with any other in-situ catalog. Moreover, the catalog does not provide the extreme ultraviolet (EUV) observations of the associated eruption or the early evolution of the CME in the coronagraphs or heliospheric imagers (HIs). 

From the in-situ perspective, the \href{https://izw1.caltech.edu/ACE/ASC/DATA/level3/icmetable2.htm#(b)}{Near-Earth Interplanetary Coronal Mass Ejections Since January 1996}\footnote{https://izw1.caltech.edu/ACE/ASC/DATA/level3/icmetable2.htm$\#$(b)} list reported by \citet{cane_2003,richardson_2010} is a complete data set of CMEs that caused associated geomagnetic storms. The catalog characterizes the in-situ parameters of the CME arrivals detected on Earth and reports candidates for the associated CME detected by the LASCO or STEREO instruments. On the other hand, a solar energetic particles (SEP) catalog of the Solar energetic particle analysis platform for the inner heliosphere \citep[\href{https://data.serpentine-h2020.eu}{SERPENTINE\footnote{https://data.serpentine-h2020.eu}}:][]{dresing_2024} data center connects SEP events detected by more than one spacecraft. This catalog provides the associated flare information, in-situ parameters, and observational details. When available, it also provides the parameters to reproduce the Graduate Cylindrical Shell model \citep[GCS:][]{thernisien_2006, thernisien_2011}, contributing crucial interplanetary information obtained by observations. Because both catalogs are focused on the in-situ perspective, they do not provide actual observations in the heliosphere or close to the Sun. 

There are several catalogs developed by the teams of different instruments that describe the remote sensing aspect of the CME evolution, such as \href{https://cdaw.gsfc.nasa.gov/CME_list/}{LASCO\footnote{https://cdaw.gsfc.nasa.gov/CME$\_$list/}} catalog \citep[][]{yashiro_2004},  {\href{https://cor1.gsfc.nasa.gov/catalog/}{COR-1\footnote{https://cor1.gsfc.nasa.gov/catalog/}}, the inner coronagraph of the SECCHI suite on board STEREO mission}, and  \href{https://www.sidc.be/cactus/}{CACTus\footnote{https://www.sidc.be/cactus/}} catalog \citep[][]{robbrecht_2004}. We can also find catalogs with detailed in-situ information like the \href{https://www.helcats-fp7.eu}{HELCATS\footnote{https://www.helcats$\-$fp7.eu}} catalog or \href{https://helioforecast.space/arrcat}{HELIO4CAsT\footnote{https://helioforecast.space/arrcat}} \citep[][]{mostl_2017}. {In the first case, the authors provide a wide set of catalogs that connect in-situ events with remote sensing observations, correlate the events with the source in the low corona, and describe different properties of the CMEs and ICMEs detected.}


\citet{hess_2017} created a comprehensive catalog of geo-effective events in solar cycle 24 using the SECCHI suite to connect each event back to its source, showing a number of statistics relating solar activity to CME properties and space weather effects. Because of the focus on geo-effective activity, this catalog only included events that reached the Earth and is no longer actively updated.

With the idea of a collaborative new generation of catalogs, we list the events observed by SoloHI from January 2022 through April 2023. We connect SoloHI observations with information available online in other catalogs in order to provide a global description of each event that is as thorough as possible. We combine observations provided by different remote sensing instruments available in the {different missions} and information provided from other catalogs. 


\section{SoloHI Contributions {in the Multi-viewpoint Analysis}} \label{SoloHIdesc} 

  \begin{figure}[h!]
   \centering
   \includegraphics[width=\hsize]{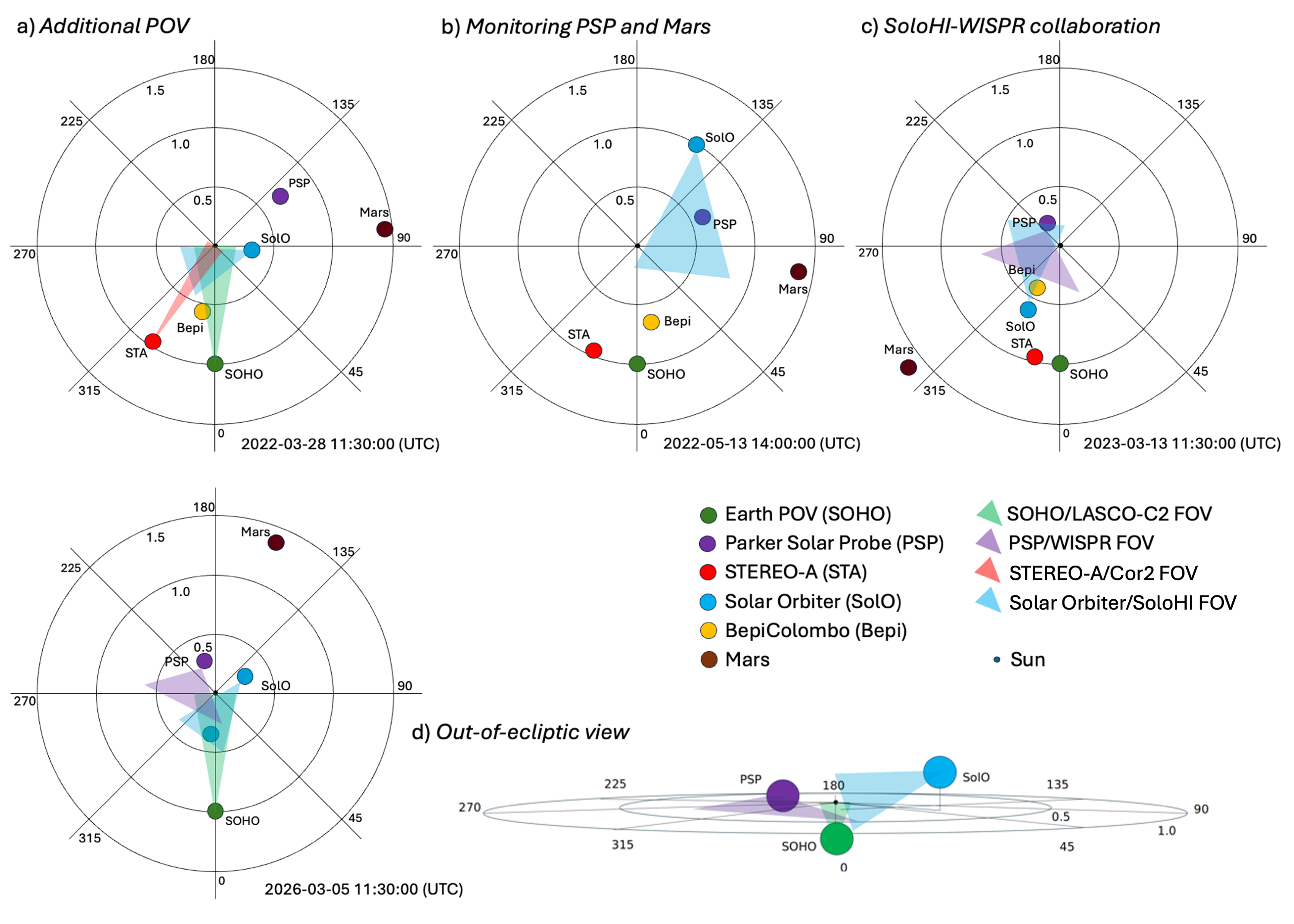}
      \caption{Illustrative sketch of the spacecraft positions and {the remote sensing instrument FOV (shaded triangle): SOHO/LASCO-C2, PSP/WISPR, STA/COR2 and SolO/SoloHI }. Upper panel: a) \textit{Additional POV:} SoloHI complements STA and near-Earth spacecraft's FOVs in multi-point observations (3 POV for Earth-directed CMEs). b) \textit{Monitoring PSP and Mars:} SoloHI Monitoring CMEs directed to Mars and PSP. The source can still be observed from Earth's POV but also from EUI on board SolO. c) \textit{SoloHI-WISPR collaboration:} SoloHI complements PSP/WISPR FOV. Both instruments can track the same CMEs with a higher resolution than previous HIs. Bottom panel: d) \textit{Out-of-ecliptic view:} SolO at 9.55$^{\circ}$ outside the ecliptic plane. We obtain a complementary FOV for CMEs detected by PSP and Earth.}\label{SolO_mission}
   \end{figure}

Solar Orbiter was launched in February 2020 and the SoloHI door was opened in June of that year. In the first 18 months, limited observations were taken during the testing and commissioning of the instrument. Since December 2021, SoloHI has been observing regularly, except for periods when the spacecraft is beyond $0.7$ au on the far side of the Sun. Figure \ref{SolO_mission} shows different observing scenarios where SoloHI has already played a key role in the multi-viewpoint analysis of CMEs during this observing period as well as the expected configuration when Solar Orbiter is due to reach a larger latitude away from the ecliptic plane. Colored circles represent the spacecraft locations at the moment of the event, and shaded triangles illustrate the {FOV of one remote sensing instrument onboard each spacecraft as reference. We show the FOVs of the C2 coronograph of LASCO (LASCO-C2) on board SOHO mission and the coronograph COR2 on board STEREO-A (STA/COR2), and the coverage of the heliospheric imagers {WISPR on board PSP and SoloHI on board SolO.}}

\textit{Additional POV:} Panel a of Figure \ref{SolO_mission} shows the complementary POV provided by SolO{/SoloHI} (light-blue circle) during the first SolO science perihelia with respect to near-Earth SOHO/LASCO-C2 (green circle) and {STA/COR2} (red circle) POVs. Since the loss of contact with STEREO-B as a secondary POV, SECCHI tools, multi-point studies, and 3D reconstruction were typically restricted to only STA and {SOHO/}LASCO observations, providing, in most cases, limited information on the CME evolution as well as larger uncertainties on all measurements. SoloHI is not only providing a third POV but also improving the analysis by observing fine-scale structures with its improved resolution at different heliocentric distances. SoloHI also provided a complementary POV for the STA/HI-1 instrument, allowing us to track the evolution of the CMEs further in time and space than only with the coronagraphs. 

\textit{Monitoring PSP and Mars:} SoloHI contributed a complementary POV for events detected at the east of the Sun, as shown in Figure \ref{SolO_mission}. During April-September 2022, SolO is more than 135$^{\circ}$ in longitude from Earth. Panel b in Figure \ref{SolO_mission} shows the FOV for SoloHI during this period (shaded light-blue triangle). It can be seen that SoloHI filled the gap in the tracking of the CME evolution that impacted Mars and PSP {in situ instruments}, contributing new information to the space weather community. With the set of RS instruments on board the SolO mission {imaging the solar disk and low corona}, these CMEs cannot only be detected but are also studied from the source to the inner heliosphere.

\textit{SoloHI-WISPR collaboration:} In this new solar mission era, where multi-viewpoint observations are key to understanding CMEs, one of the most exciting collaborations is between {SoloHI and PSP/WISPR}. WISPR is the only remote sensing instrument on board PSP and is always pointing at the west of the Sun with respect to PSP with a FOV of 95$^{\circ}$ of elongation. With the similarities between the instruments and the high-resolution observations, this joint effort can bring to the community invaluable information about the internal structures of CMEs and other structures. Panel c of Figure \ref{SolO_mission} shows the potential collaboration between SoloHI (light-blue) and WISPR (violet), where both instruments present a complementary POV, allowing to observe CMEs from opposite sides and the ability to study their evolution with a resolution that has never before been possible.

\textit{Out-of-ecliptic view:} In the upcoming years of the mission, SolO will reach continuously larger latitudes outside of the ecliptic plane. SoloHI will provide a unique FOV, as shown in panel d in Figure \ref{SolO_mission}. This scenario will give us new constraints in understanding the evolution of CMEs by providing a new viewpoint from a significantly different plane than those ever provided by previous generations of remote sensing instruments. This configuration will allow us to increase our knowledge about the 3D structure and the characteristics of CMEs that originate farther from the ecliptic, closer to polar coronal holes (CHs).

\section{SoloHI catalog}
\label{SoloHIcat}

The comprehensive list of the events observed by SoloHI is found here on the \href{https://science.gsfc.nasa.gov/lassos/ICME_catalogs/solohi-catalog.shtml}{website\footnote{https://science.gsfc.nasa.gov/lassos/ICME$\_$catalogs/solohi$\-$catalog.shtml}}. We show a subsection of the SoloHI events in Appendix~\ref{Appendix} Table \ref{table:EventsRSW} for the first two science perihelia in 2022 as an example to demonstrate the main information provided by the SoloHI catalog. A condensed list of all the 140 detected events during the period of SoloHI observations up to April 2023 is provided in Table \ref{ListSoloHI} in Appendix~\ref{Appendix} for the reader's quick reference. In this section, we provide a comprehensive overview of all the information collected for the catalog, including the resources available online.

SoloHI data can be found in the Solar Orbiter Archive (\href{http://soar.esac.esa.int/soar/}{SOAR\footnote{http://soar.esac.esa.int/soar/}}) and on the \href{https://solohi.nrl.navy.mil}{SoloHI webpage\footnote{https://solohi.nrl.navy.mil}}. On {SoloHI} webpage, the reader can also find a step-by-step tutorial for downloading, processing, and creating movies of the SoloHI images with IDL routines. To request Python routines, the reader may contact the SoloHI team. 

{To populate the catalog, we identify the events by performing a visual inspection of any dynamic structure in the SoloHI movies. The selected structures exhibit a distinct density enhancement that can be tracked over time. Notably, many of these structures deviate from the well-behaved CME, as they may be partially detected (e.g., when a CME propagates toward SolO, only its eastern portion may be visible in the SoloHI FOV). Furthermore, these structures often display significant distortion observing them from the heliospheric imager perspective or even faint or poor signal. To ensure accuracy in the selection, the final list of CMEs presented in the catalog is the result of a cross-checking analysis of the presence of the CME in the multi-instrumental dataset described in the following subsections.}

Once the CME is identified, we note its first appearance {and use it as an event indicator in the catalog. Each event is labeled then by the start date in the format \textit{YYYY MM/DD HH:MM}.} Then, we conduct a morphological classification depending on the signatures observed in SoloHI observations, which is presented in Section \ref{Morpho}.


Once we have the event identified, we look at the configuration of {multiple spacecraft} at the time of the observation. For this, we use the open-source tool \href{https://solar-mach.streamlit.app/?embedded=true}{Solar-MACH\footnote{https://solar$\-$mach.streamlit.app/?embedded=true}} \citep{gieseler_2023} available online. This tool allows us to visualize not only the position of spacecraft or planets in the heliosphere but also the solar magnetic connection with a nominal Parker spiral. We identify the position of SolO for each listed event. Since we know where the SoloHI FOV is pointing, we can use SoloHI observations to estimate the CME propagation angle. We also look at the position of other spacecraft to determine which ones could observe the same event for both the remote and in-situ instruments.

\subsection{Coronagraphs and heliospheric imager detection}

Our next step is to identify all the information provided by the available coronagraphs and heliospheric imagers. We thoroughly examine the data obtained by COR1-2 on board STA and LASCO/C2-C3 coronagraphs on board SOHO using the \href{https://cdaw.gsfc.nasa.gov/movie/}{CDAW Data Center\footnote{https://cdaw.gsfc.nasa.gov/movie/}}. Once the event has been visually identified in the FOV of different instruments, we link the corresponding event movies with the catalog. We report if the particular event was also detected by WISPR or not. We perform a visual inspection of the movies available for each encounter in the \href{https://wispr.nrl.navy.mil/encounter-summaries}{WISPR webpage\footnote{https://wispr.nrl.navy.mil/encounter-summaries}}.

\subsection{Solar source identification}

Once we identify the event in the FOV of a given coronagraph, we track its trajectory back to the solar surface and connect the event with the source region. We use {data from the EUV instruments Extreme Ultra-Violet Imager \citep[EUVI:][]{wuelser_2004} part of the SECCHI suite on board STA, the Atmospheric Imaging Assembly \citep[AIA:][]{lemen_2012} on board SDO,} and the {data provided by the Extreme Ultraviolet Imager \citep[EUI:][]{rochus_2020} on board} SolO, to identify the activity that leads to the detected CME. We perform a visual inspection of the EUV observations and identify the region where the eruption takes place. We use the movies available on the CDAW Data Center in two different wavelengths (193$\AA$ for AIA and 195$\AA$ for EUVI and 304$\AA$ for both) and link them to the SoloHI catalog. {For EUI, we use the database provided by the Solar Influences Data Analysis Center \href{https://www.sidc.be/EUI/data/}{(SIDC\footnote{https://www.sidc.be/EUI/data/})}}. When possible, we report the active region number, its location, associated filament material or flare, and the type of the flare. For the entire description of the source, we also link the catalog with the \href{https://solarmonitor.org}{Solar Monitor\footnote{https://solarmonitor.org}} webpage with the information {of the magnetogram, observed on the day of the event, provided by the Helioseismic Magnetic Imager \citep[HMI:][]{scherrer_2012} on board SDO mission.}


\subsection{Connecting to Operations}

All the collected information is compared with the CME catalog made by the Moon to Mars (M2M) Space Weather Analysis Office. They perform a daily report of CME that may or may not impact other spacecraft and provide this information on \href{https://ccmc.gsfc.nasa.gov/tools/DONKI/}{DONKI\footnote{https://ccmc.gsfc.nasa.gov/tools/DONKI/}} catalog, developed by CCMC. They provide a vast amount of complementary information regarding the possible trajectory of the event and the WSA-ENLIL+CONE model \citep{arge_2000} that helps to predict the time of impact of the CMEs on other spacecraft or planets. We link our catalog to the description made by the M2M team.

\subsection{In-situ detections}

From the predicted impact information provided by the M2M catalog, we look at the {magnetic and plasma properties data detected in-situ and} provided by the different instruments on board WIND, SolO, PSP, and STA spacecraft. We provide plots with the magnetic field information (total magnetic field and magnetic field components) and plasma properties (density, temperature, velocity, and plasma beta) during the time range of the probable impact. 

\subsection{Other Catalogs}

We also link each of the events with the \href{https://cdaw.gsfc.nasa.gov/CME_list/}{SOHO LASCO CME catalog\footnote{https://cdaw.gsfc.nasa.gov/CME$\_$list/}} \citep{yashiro_2004} to complement our information with their reports. Another important contribution linked to the catalog is the information collected in the catalog based on STA/HI observations HICAT \citep{harrison_2018} under the umbrella of the Heliospheric Cataloguing Analysis and Techniques Service (HELCATS) project. This catalog reports CMEs observed by the inner telescope of the STA/HI (HI-1) and provides information on CME dates and properties that complement the SoloHI observations. We also look for the flare information in the \href{https://xrt.cfa.harvard.edu/flare_catalog/}{XRT\footnote{https://xrt.cfa.harvard.edu/flare$\_$catalog/}} flare catalog developed by \citet{watanabe_2012}. This flare catalog collects flares detected during the Hinode mission \citep{kosugi_2007}, and is linked to the Heliophysics Events Knowledgebase (\href{https://www.lmsal.com/hek/index.html}{HEK\footnote{https://www.lmsal.com/hek/index.html}}) database that provides all basic information regarding the associated flare eruption \citep{timmons_2023}.

\subsection{Early SoloHI Observations in 2022}

Table \ref{table:EventsRSW} in Appendix \ref{Appendix} provides the details of a subsection of the catalog with 27 events detected during the first two science perihelia, in the date range of 03 March to 05 April 2022 and 08 October to 01 November 2022. During this period, the cadence of the instrument is higher than the rest of the orbit, because it is closer to the Sun. We provide the complementary information obtained from other missions that are available online and linked in the catalog. The hyperlinks are available on the catalog webpage.

Out of the 27 events identified, we found 16 events during the March-April period of 2022 and 11 events during October 2022. Most of these events were also detected by other remote sensing instruments onboard other spacecraft: 25 events were seen by SOHO/LASCO (C2 and C3) and STA/COR2. We did not identify any event detected by SoloHI and WISPR simultaneously during this period, as the orbits of the two spacecraft were not well aligned. We identified 16 source regions associated with events on the solar surface, mostly during the first science perihelia, where the FOV of SoloHI overlapped with the FOV of Earth-based and near-Earth observations. Nine events were detected by at least one in-situ instrument. Out of those events, 3 were detected by more than one spacecraft simultaneously.

\begin{figure}[h!]
   \centering
   \includegraphics[width=\hsize]{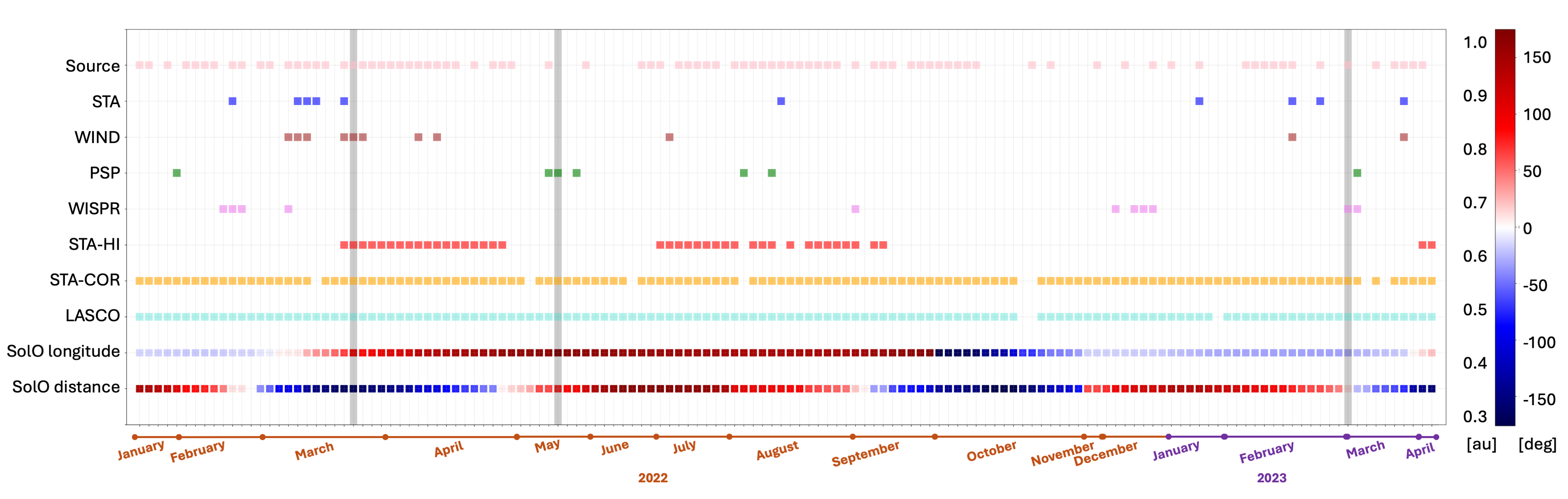}
      \caption{{Graphical overview of the information collected for each event listed in the catalog during the analyzed period. Pink squares represent whether the EUV source was identified. LASCO, STA/COR, STA/HI, PSP/WISPR represent the remote observations (cyan, orange, red, and violet square, respectively), and PSP, WIND, STA if the event was detected in-situ (green, brown, and blue square, respectively). SolO distance and longitude provided for the location of the spacecraft at the moment of the observation in the SoloHI FOV (color bar on the right). The x-axis is ordered by the date of each detection, so months with more detected events are larger.  Grey vertical lines highlight the three examples described in the next Section \ref{SoloHIdesc}.}}
      \label{OverCat}
   \end{figure}

{Figure \ref{OverCat} provides a graphic overview of the catalog coverage for the 140 events listed in Table \ref{ListSoloHI}. Pink squares at the top denote if an EUV source region was identified. Colored squares in the next row indicate if an in-situ detection was identified at STA, WIND, or PSP. The next rows indicate remote sensing observations in WISPR, STA-HI1, STA-COR2, and LASCO. The color of the boxes in the bottom rows represents the heliocentric distance and HEE longitude of SolO at around the time of the first observation in the SoloHI FOV. The x-axis of Figure \ref{OverCat} is the date of each individual event detected for the catalog, which is why the months have a non-uniform size. Vertical gray lines highlight the specific examples discussed in the next Section \ref{SoloHIdesc}. Notably, the scenarios where SoloHI is monitoring PSP or collaborating with WISPR are sporadic but recurring throughout the analyzed period. Conversely, SoloHI frequently provides a third viewpoint and complements STA-HI observations for numerous 2022 events.}
 
\section{Morphological classification of SoloHI events}
\label{Morpho}

In this section, we perform a general analysis of the dataset of 140 events detected by SoloHI from January 2022 through April 2023. 

\begin{figure}[h!]
   \centering
   \includegraphics[width=\hsize]{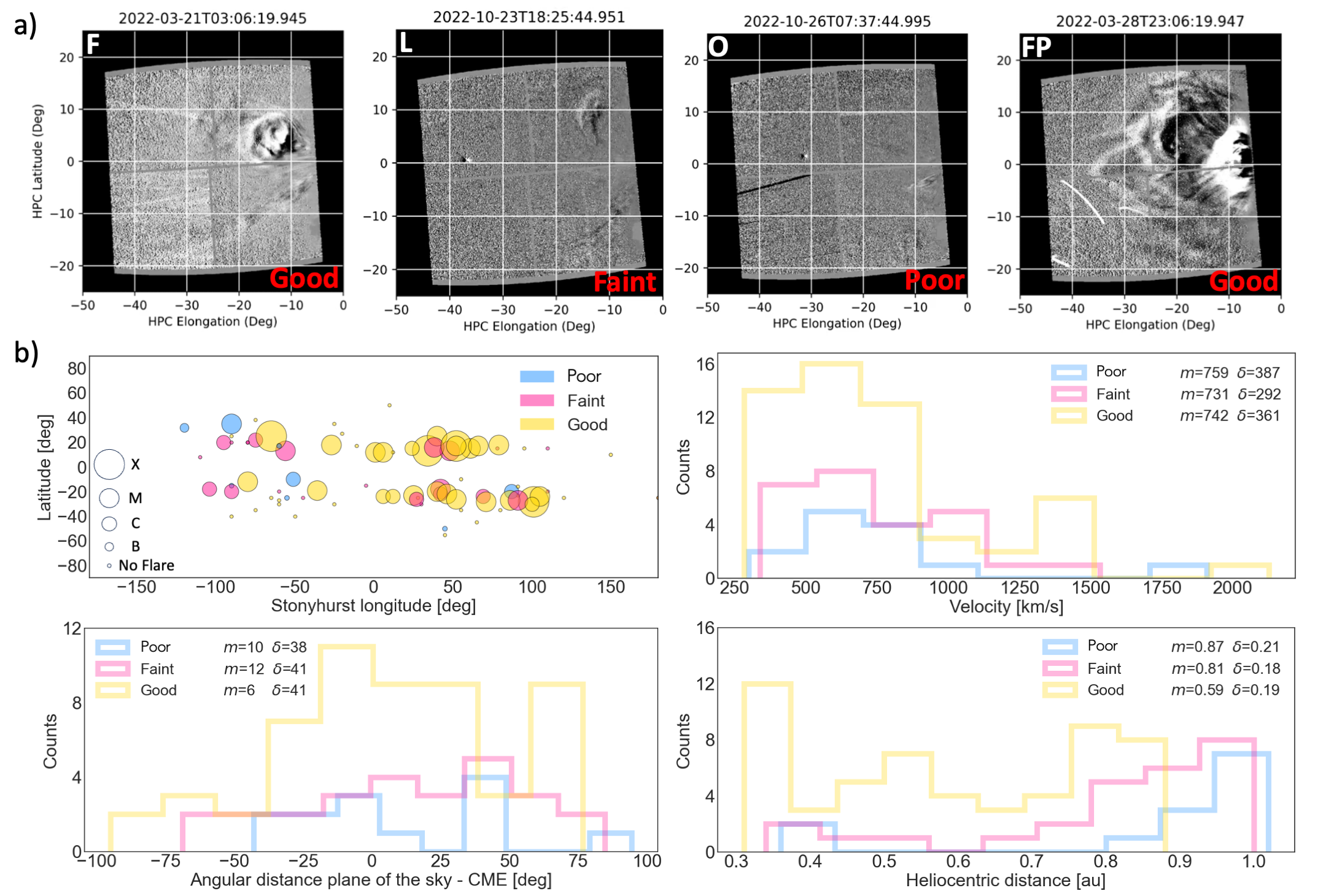}
      \caption{Panel a: Examples of different morphological classifications of CME in the SoloHI observations denoted by white letters. The label \emph{F} (Flux Rope) denotes a well-identified 3-part structure CME. \emph{L} (Loop) corresponds to a CME observed from another perspective. \emph{O} refers to other structures that cannot be well identified, and \emph{P} (Preceding) is added to the other classification if there is a CME evolving while we detect a new CME in the SoloHI FOV. In red, we denote the quality classification made for the intensity of the events observed in SoloHI FOV: Good, Faint, and {Poor}. Panel b: We show the location of the identified source in a latitude-longitude map in the upper-left panel. Colors indicate the quality classification of the event, and circle sizes indicate the flare classification. We also present histograms for the radial velocity reported in the DONKI catalog (upper-right), the angular distance between the image plane for SoloHI and the CME longitude propagation reported by the DONKI catalog (bottom-left) and the distribution of the events as a function of the heliocentric distance of SolO at the moment of the event detection (bottom-right).
              }
         \label{types}
   \end{figure}

As a first step, we label the CMEs depending on their morphological signatures that we can identify in SoloHI movies. We use some of the labels described in the multi-viewpoint catalog of events detected by STEREO-A and STEREO-B simultaneously \citep{vourlidas_2017}. Panel a of Figure \ref{types} shows examples of different morphologies of the CME that we identified and their assigned labels. We categorize a CME as \emph{F} (Flux Rope) if it exhibits the 3-part CME structure (core-cavity-front). We label the CME with \emph{L} (loop) if it is observed from the perspective that cannot identify the cavity or the core. We assign an \emph{O} (Other) to any CME that cannot be identified as either F or L. Finally, we assign \emph{P} (Preceding) to a CME if there are two simultaneous CMEs in the SoloHI FOV. This classification is only based on the features observed from the SoloHI POV. Complementary POVs from other instruments may exhibit the features of a different classification. 

Table \ref{tab:statistics} shows a summary of the classifications made for the 140 events detected. We see that more than half of the events were classified as F, while almost 20$\%$ of the events are classified as 'O's, as no clear structure could be identified in SoloHI. We could relate more than half of the events to a corresponding source region, 22 of those sources were on the limb or on the backside from the Earth's perspective. Out of those 22 sources, 7 were identified with STA/EUVI data, and 15 were identified with SolO/EUI. The number of flares associated with the eruptions is 41, and the number of filaments is 46 out of the total 140 events. 


\begin{table}
\caption{Classification of the event detected by SoloHI from January 2022 until April 2023. Column 1 shows the data time periods that are divided by the SolO heliocentric distance (R) in au (Far: R $\geq$ 0.7; Mid: 0.7 $>$ R $>$ 0.4; Close: 0.4 $\geq$ R). Columns 2, 3, and 4 indicate the morphological classification of the events detected by SoloHI: Flux rope (F), Loop (L), and Other (O), respectively. Column 5 gives the percentage of the total number of CMEs detected. Column 6 gives the number of events with identified source regions. Columns 7 and 8 give the number of source regions with flares or filaments related to the eruptions, respectively.}
\label{tab:statistics}
    \begin{tabular}{l | c c c | c | c c c}


        Heliocentric Distance [au] & F & L & O & $\%$ & So & Fl & Fi \\
        \hline
        R $\geq$ 0.7 (Far) & 37 & 26 & 12 & 54 & 42 & 24 & 21 \\
        0.7 $>$ R $>$ 0.4 (Mid) & 24 & 7 & 7 & 27 & 27 & 6 & 12 \\
        0.4 $\geq$ R (Close) & 11 & 7 & 9 & 19 & 19 & 11 & 13 \\

        \hline
        Total & 72 & 40 & 28 &  & 88 & 41 & 46 \\ 
        $\%$ & 51 & 29 & 20 &  & 63 & 29 & 33 
    \end{tabular}
\end{table}

We also perform a quality classification depending on the intensity of the event in the SoloHI FOV. During this period, we found 73 good events that were tracked through the entire SoloHI FOV. The rest of the events were classified as faint events (42), where we can identify structures, but the observation faded in the SoloHI FOV over time, and {poor} events (25), where structures are difficult to detect without enhancing the images with additional processing. Examples of these quality classifications are shown in panel a of Figure \ref{types}.

Panel b of Figure \ref{types} shows the distribution of the identified source regions and the quality classification of the corresponding events detected in the SoloHI movies (upper-left). The circles correspond to the source location. Yellow circles represent the source of good-quality events. Pink and light-blue circles represent the faint and {poor} detection in the SoloHI observations, respectively. In latitude, we can see two active region (AR) populations centered around $\pm25\,$deg, as is expected for this period of the solar cycle. The size of the circles represents the classification for the flare associated with the eruption. We reported four X-class flares, 23 M-class flares, 13 C-class flares, and one flare class B. We did not find an associated flare in 47 identified sources. In those events, we find a slight tendency towards {poor} events, which correlate with the weaker flare classification, and the higher classification (M and X-class flare) is mainly present for good-quality events.


In the upper-right panel, we present the distribution of the CME radial velocity reported by the DONKI catalog. This velocity is computed with the CME Analysis Tool (CAT) software system used at the NOAA Space Weather Prediction Center (SWPC) \href{https://ccmc.gsfc.nasa.gov/tools/SWPC-CAT/}{SWPC$\_$CAT\footnote{https://ccmc.gsfc.nasa.gov/tools/SWPC-CAT/}}. To calculate the radial velocity, they initially used the CAT model to obtain the 3D geometry of the CME using observations from SOHO/LASCO-C2/3 and STA/COR2 and taking into account the position of those spacecraft. They adjust the propagation angle, angular width, and leading-edge distance, using the latter to generate a position-time plot. By performing a linear fit on this plot, they obtain the radial velocity of the modeled leading edge, which is subsequently reported in the DONKI catalog. The majority of the events have velocities smaller than 1000$\,$km$/$s. The average CME speed computed is $\approx$741$\,$km$/$s. In contrast, \citet{yashiro_2004} reported an average projected speed of 300–500$\,$km$/$s, with a maximum speed of approximately 2600 km/s, based on their analysis of 6599 CMEs in the SOHO/LASCO CME catalog. The discrepancy between these average speeds may be attributed to projection effects present when computing speeds with a single instrument. No clear correlation is identified between the velocity and the quality classification of the events in the SoloHI FOV. 

The bottom-left histogram of panel b in Figure \ref{types} shows the distribution of the angular separation between the image plane for SoloHI and the longitude propagation of the CME also obtained with the DONKI catalog. Negative angles represent the CME propagating toward SolO, while positive angles denote CMEs that propagate further than the image plane. We observe that good events tend to cluster within $\pm45\,$deg of the image plane, whereas faint and {poor} events are more dispersed. Additionally, we detect an asymmetry in the distribution of larger angles among good events, with a greater number of detected events propagating beyond the image plane than in front of it. Similar behavior can be observed for the other two event classifications.  
While it is expected for this quantity to be directly associated with the intensity of the detection, \citet{howard_2012} performed a study to demonstrate that HIs have the sensitivity and capability to detect CME features in a wide range of CME propagation angles and not only those close to the image plane. Thus the SoloHI observations are consistent with the results of \citet{howard_2012}. 

Finally, the bottom-right panel of Figure \ref{types} shows the distribution of the SolO heliocentric distance at the moment of the detection of the events. We observe that faint and {poor} events correlate with larger heliocentric distances, as is expected, while good events are more dispersed.



\section{Cases of study: Examples of the capabilities of the catalog}
\label{CasesOfStudy}

To summarize the coverage of this catalog, in this section, we show three CMEs detected during the particular scenarios listed in Section \ref{SoloHIdesc}: (a) Additional POV, (b) Monitoring PSP and Mars, and (c) SoloHI-WISPR collaboration. Scenario (d) is not available at the moment of this publication. We expect to have first observations out-of-ecliptic in the upcoming years.

Here, we describe events observed in each of the scenarios utilizing the information present in the catalog.

\subsection{Additional POV}

   \begin{figure}
   \centering
   \includegraphics[width=\hsize]{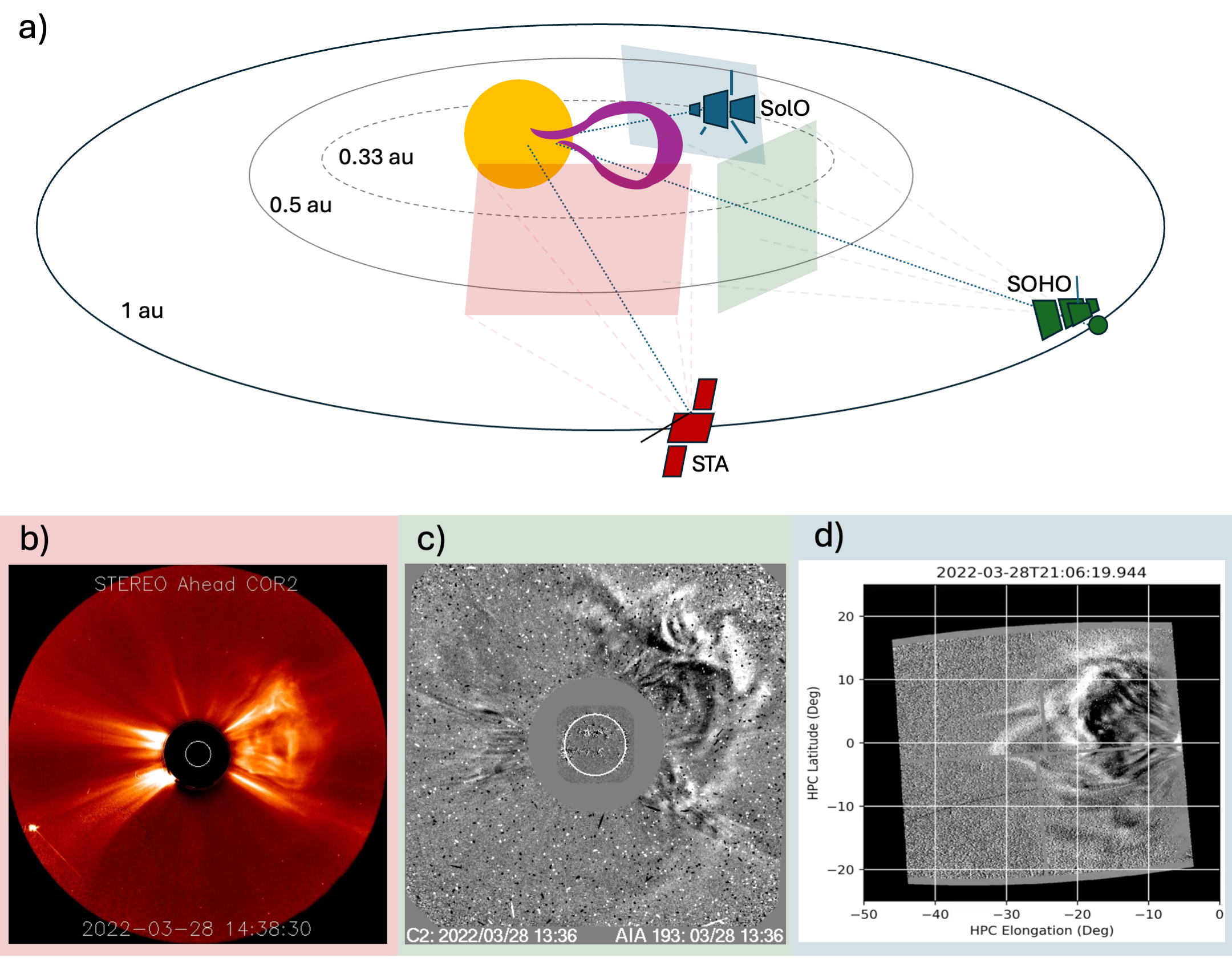}
      \caption{Panel a) Illustrative sketch highlighting the POV of each spacecraft that observed the March 28th event: POV of SolO is in blue, SOHO is in green, and STA is in red. A cartoon representation of the CME direction is given in magenta. Panel b) STA/COR2 observation of the March 28 event at 14:38:30$\,$UT. Panel c) SOHO/LASCO-C2 observations of the same event at 13:36$\,$UT. Panel d) SolO/SoloHI observation of the CME detected at 21:06:19$\,$UT. Orange and violet arrows denote the same structures observed by all the instruments.
              }
         \label{20220328_sc}
   \end{figure}

Panel a of Figure \ref{20220328_sc} shows a sketch of the spacecraft configuration on 28 March 2022. STA is at 0.97$\,$au from the Sun and -33.2$^{\circ}$ from the Earth, and SolO is at 0.33$\,$au and 84$^{\circ}$ from the Earth. The bottom panels in Figure \ref{20220328_sc} show the observations of a CME during this spacecraft configuration. 

Panel b in Figure \ref{20220328_sc} shows the CME detected in the FOV of COR2 coronagraph on board STA. The first appearance of the CME in COR1 is at 11:36$\,$UT, and at 12:08$\,$UT in COR2. In both cases, we observe the CME's left side propagating towards Earth. HICAT catalog reports the first appearance of the event in HI-1A at 14:08$\,$UT on 28 March 2022.

The CME makes its first appearance in LASCO-C2 at 12:00:07$\,$UT and LASCO-C3 at 13:30:37$\,$UT. Panel c in Figure \ref{20220328_sc} shows the CME in LASCO-C2. As the event is directed toward Earth, CDAW identifies it as a halo CME with the source slightly shifted to the North-West. Halo CMEs are crucial to understand the impacts of Solar activity on the Earth's environment. However, this viewpoint does not provide quality observation for the understanding of the morphology and evolution of the CME. 

Panel d shows the SoloHI observation at 21:06:19$\,$UT on 28 March 2022. We detected the shock of the CME around 12:42:19$\,$UT and the leading edge at 13:30:19$\,$UT in the SoloHI FOV. The CME was cataloged as an F type in the catalog. This event is described in \citet{hess_2023}, where they compare the structures detected in the SoloHI images with the observations provided by HI-1A on board the STEREO mission. 

Since the loss of STEREO-B as a complement for STA, describing CMEs with a 3 POV quadrature has been impossible. While LASCO-C2 and C3 provided the Earth-directed POV, both COR1, and COR2 are centered on the Sun and detect CMEs evolving on both sides, and HI-1 and HI-2 are focused on the evolution at the west of the Sun. This configuration is complemented by the SoloHI FOV that points to the east of the Sun with respect to the SolO spacecraft's location. This is a unique configuration that allows tracking CMEs towards Earth from three different POVs. 
SoloHI contributes as a key instrument acting as a complementary FOV in the task of describing Earth-directed CMEs. With the advantage of SoloHI observations as an additional POV to the existing regular observations from the LASCO and STA, the community can perform more constrained three-dimensional CME reconstruction of the flux rope using, for example, the GCS model, which improves its accuracy by using multiple POV to describe a particular event \citep{thernisien_2011b}.

The \href{https://cdaw.gsfc.nasa.gov/movie/make_javamovie.php?stime=20220328_1000&etime=20220328_2000&img1=lasc2rdf&img2=lasc3rdf}{LASCO C2-C3\footnote{https://cdaw.gsfc.nasa.gov/movie/make$\_$javamovie.php?stime=20220328$\_$1000$\&$etime= \\ 20220328$\_$2000$\&$img1=lasc2rdf$\&$img2=lasc3rdf}} and \href{https://cdaw.gsfc.nasa.gov/movie/make_javamovie.php?stime=20220328_1000&etime=20220328_2000&img1=sta_cor1&img2=sta_cor2}{COR1-2\footnote{https://cdaw.gsfc.nasa.gov/movie/make$\_$javamovie.php?stime=20220328$\_$1000$\&$etime= \\ 20220328$\_$2000$\&$img1=sta$\_$cor1$\&$img2=sta$\_$cor2}} movies for this event can also be found in the SoloHI catalog. We aim to analyze the evolution of the CME and the complexity of its AR in a forthcoming article.

\subsection{Monitoring PSP and Mars}

The combination of remote sensing observations with in-situ monitoring can provide valuable insight into the structure and evolution of CMEs and their ability to drive space weather throughout the solar system. SoloHI can contribute to these observations by providing high-resolution, high-cadence images from the unique and changing perspective of the SolO orbit. A particular scenario that highlights the monitoring by both PSP and the Mars Atmosphere and Volatile Evolution (MAVEN) Mission \citep{jakosky_2015} at Mars is on May 13, 2022. SoloHI captured images of the CME on May 12, 2023, en route to the in-situ monitors. This type of configuration is not only important for the possibility of studying the evolution of CMEs more globally with the various in-situ monitors, but also provides additional monitoring of Mars, enabling a more complete study of space weather at locations other than the Earth.

    \begin{figure}
   \centering
   \includegraphics[width=\hsize]{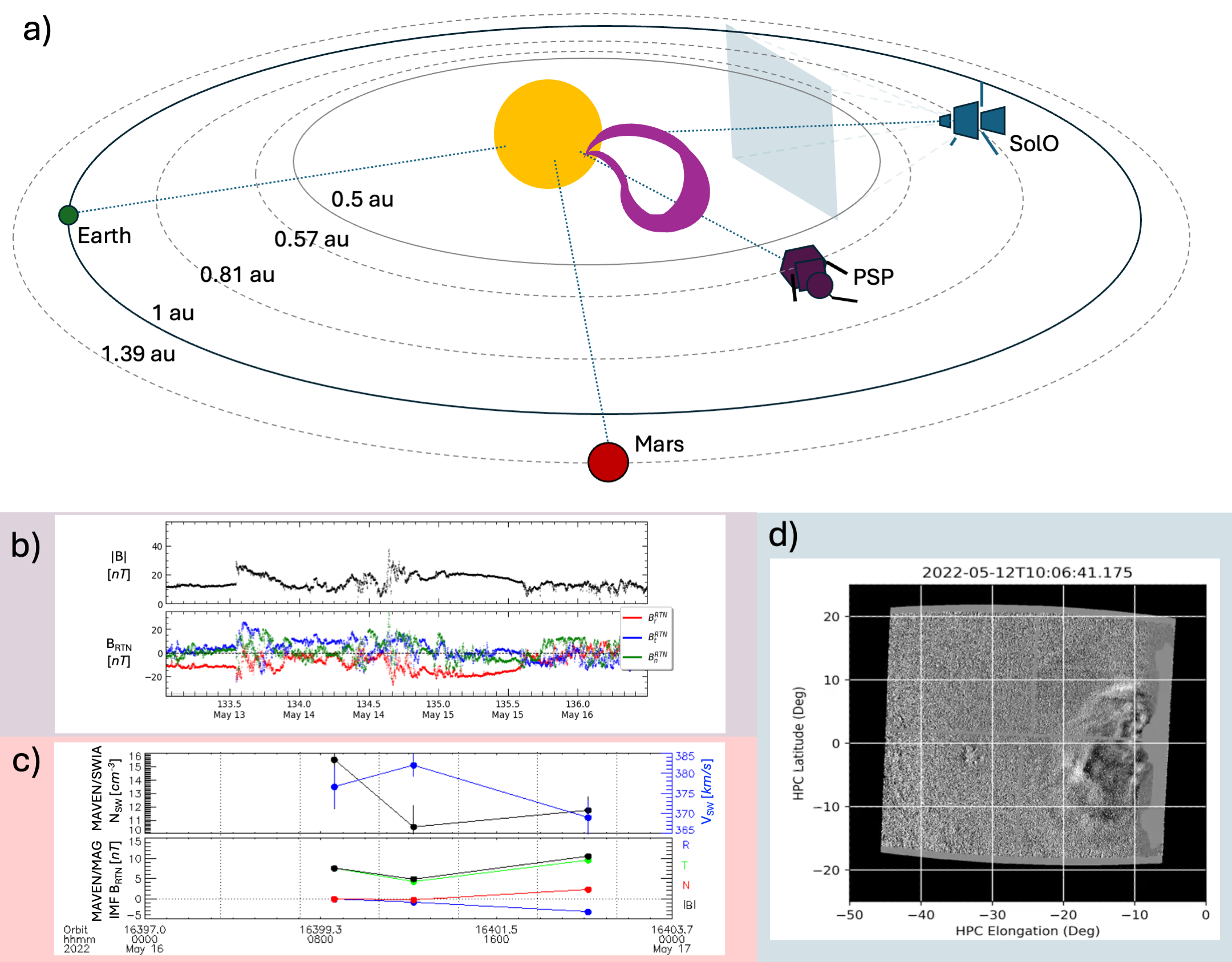}
      \caption{Panel a) Same as panel a in Figure \ref{20220328_sc} but for May 13, 2022, with SolO in blue, PSP in purple, and Mars in red. Panel b) PSP in-situ signatures detected by FIELDS in the period of May 13 - 17, 2022. The upper panel presents the total magnetic field, and the lower panel shows the magnetic field component. We can detect the arrival of two different structures denoted by the increase of the total magnetic field, the first one on May 13, 2022, at -13:06$\,$UT, and the second on May 14 at -15:13$\,$UT. Panel c) in-situ signatures of the density, velocity, and magnetic field detected by MAVEN, monitoring Mars on May 16 - 17, 2022. A clear perturbation of the velocity and density can be detected $\sim$12:00$\,$UT on May 16, 2022. Panel d) SolO/SoloHI observation detected at 10:06:41$\,$UT on May 12, 2022.
              }
         \label{May5_RS}
   \end{figure}

On May 11, 2022, at 22:08:41$\,$UT SoloHI detected the first appearance of a CME directed towards PSP. This CME is classified as an FP in the catalog, which means that we observe a flux rope with a preceding CME in the SoloHI FOV (panel d in Figure \ref{May5_RS}). A second CME appears in the SoloHI FOV on May 13, 2022, at 14:08:41$\,$UT. Classified as an F morphology, this CME is also directed towards PSP. Panel a in Figure \ref{May5_RS} shows a sketch of the spacecraft and planet configuration at the moment of the detection of the CMEs. 

The WSA-ENLIL+cone model run, found in the CME M2M catalog, predicts the impact of the first CME with PSP on May 13, 2022, at 00:35$\,$UT and with Mars on May 15, 2022, at 18:08$\,$UT. For the second CME, the model predicted the arrival at PSP on May 14, 2022, at 23:41$\,$UT, and at Mars on May 17, 2022, at 11:53$\,$UT. On panel b of Figure \ref{May5_RS}, PSP/FIELDS detection is presented for May 13 - 17, 2022. We detected the first shock arrival around 13:00$\,$UT on May 13, 2022. The magnetic obstacle cannot be clearly identified for this CME, and the plasma detection is not conclusive. A second arrival can be detected on May 14, 2022, at $\sim$15:00$\,$UT. This second CME has a purely radial structure that can be related to a leg of the CME. Plasma measurements for this period are not available. The arrival time does not match exactly with the predictions made by the WSA-ENLIL+cone model. During this period, a lot of activity with the coronagraphs was detected, and that can add complexity to the interpretations of in-situ data.

Panel c on Figure \ref{May5_RS} shows an arrival detected at MAVEN. The upper panel presents the variation of the plasma density and velocity, while the bottom panel shows the variation of the total magnetic field and its components. The data is low-resolution for this time period, but it suggests an increase in the velocity and a drop in the density that can be detected on May 16, 2022, around 08:40$\,$UT. This change in the plasma may be related to the arrival of the second CME also detected by PSP.

SoloHI data enables a kinematic study of the CME front by allowing for the determination of the CME velocity, trajectory, and acceleration. This can help to reduce the error in the predicted arrival times. In an upcoming article, we will conduct a detailed analysis of this event, focusing on its specific characteristics and underlying mechanisms. 

\subsection{SoloHI-WISPR collaboration}

\begin{figure}
   \centering
   \includegraphics[width=\hsize]{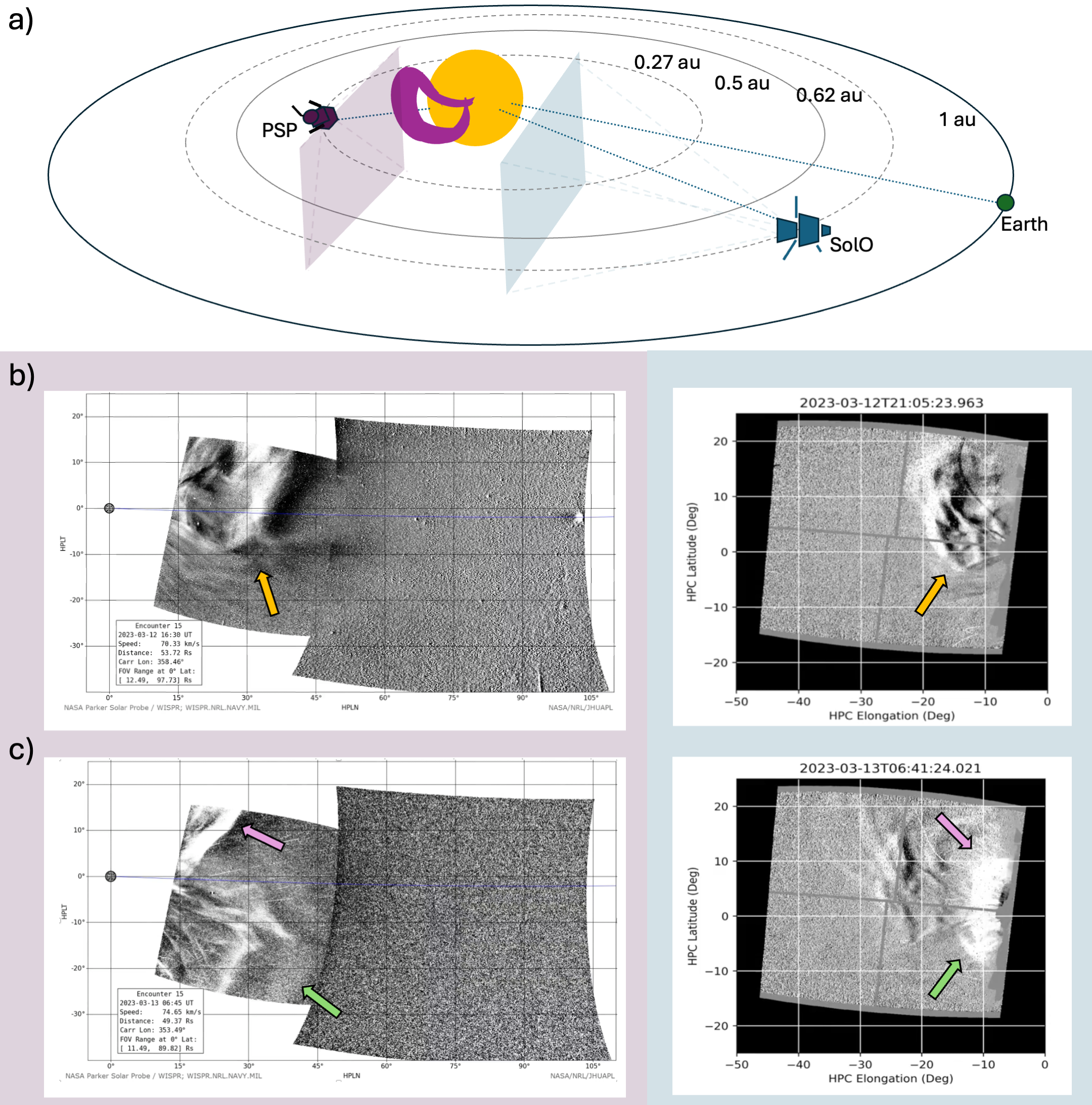}
      \caption{Panel a) Same as panel a in Figure \ref{20220328_sc} but for March 12, 2023, with SolO in blue and PSP in purple. Panel b) PSP/WISPR (left) and SolO/SoloHI (right) observations for the CME detected on March 12, 2023, at 17:15:00$\,$UT and 21:05:23$\,$UT, respectively. Orange arrows denote the same structures observed by both instruments. Panel c) PSP/WISPR (left) and SolO/SoloHI (right) observations for two CMEs detected on March 13, 2023, at 06:45:00$\,$UT and 06:41:24$\,$UT, respectively. The green arrow denotes the leading edge and the low-density cavity for the first CME in both instruments. The pink arrow indicates the second CME.
              }
         \label{SoloHI-WISPR_sketch}
   \end{figure}

SoloHI observes the heliosphere at the east of the Sun while WISPR points to the west with respect to the PSP and SolO orbits. This allows for the viewing of CMEs from opposing viewpoints when WISPR is observing during a PSP encounter within $0.25$ au, and the two spacecraft are roughly between $90^\circ$ and $180^\circ$ of longitude from each other. To get the benefit of the high-resolution images from SoloHI, SolO should be near perihelion as well. In the periods covered by the catalog to date, the PSP and SolO orbits were oppositely aligned, meaning they were not approaching perihelion at the same time, and typically, only one of the HI instruments had available high-resolution images, thus this configuration still has limited cases. Out of the 140 events reported in this paper, we identified three well-observed events and three faint events that were detected by both instruments. Figure \ref{SoloHI-WISPR_sketch} shows the well-observed events detected on March 12 - 13, 2023. Panel a of Figure \ref{SoloHI-WISPR_sketch} shows a sketch of the position of SolO and PSP in those joint observations. 

The first major CME was detected simultaneously by both instruments on March 12, 2023. At the moment of the detection, Solar Orbiter was at 0.88$\,$au and -18.2$^{\circ}$ relative to the Earth and PSP at 0.68$\,$au and -149.6$^{\circ}$ on longitude from the Earth. Panel b of Figure \ref{SoloHI-WISPR_sketch} shows the observations made by WISPR (left) and SoloHI (right) at 17:15$\,$UT and 21:05:23$\,$UT, respectively. 
Although the WISPR instrument did not detect the first appearance of the CME in its FOV as it was not observing at the time, we can identify different structures of the CME in its later evolution. The CME presents a slightly distorted shock with a leading edge flattened on the side, followed by a concave structure (yellow arrow in Figure \ref{SoloHI-WISPR_sketch}).

SoloHI detected the first appearance of the March 12th CME at 07:29:48$\,$UT. This CME is classified as an F in the SoloHI catalog. In the observations, we identified the same concave structure detected in WISPR observations and the leading edge (orange arrow in Figure \ref{SoloHI-WISPR_sketch}). 

Two other CMEs were observed in the FOV of WISPR on March 13, 2023, as shown in panel c of Figure \ref{SoloHI-WISPR_sketch}. The first CME is detected at 04:45$\,$UT and impacts PSP. A clear cavity is detected for this CME (green arrow in Figure \ref{SoloHI-WISPR_sketch}) and can be followed in almost the entire WISPR FOV. The second CME shock (violet arrow) is partially detected at 05:45$\,$UT in the north part of the FOV. We detect both CMEs arriving at the same time in the SoloHI FOV (05:05:24$\,$UT). While in the upper part of the CME, we can only detect a shock, in the bottom part, the cavity of the first CME is identified. All the mentioned features are denoted with arrows in Figure \ref{SoloHI-WISPR_sketch}.

Fortunately, more recent periods have seen better alignment between SolO and PSP, and we anticipate more events that allow us to combine images from these instruments to better resolve CME features in the heliosphere.

\section{Summary and conclusions}
\label{Conc}

{Understanding CME internal and external structures and their evolution in the heliosphere remains an active area of research. Previous generations of solar missions provided data regularly, spanning over more than two solar cycles and enabled significant advancements in our understanding of CME global structures. The new generation of Solar missions such as PSP, SolO are encounter missions that provide high-resolution data during brief periods of close solar approach. These missions provide unprecedented imaging of the CME internal structures and interactions with coronal structures and solar wind. We anticipate that by combining these new observations with the contextual observations from synoptic mission, the scientific community will be able to make new advancements in CME evolution and propagation. }

{Unique among the new generation of encounter missions,} SolO has a full suite of instruments that contribute with different measurements and observations. SoloHI, in particular, is a key instrument to incorporate in the study of the evolution of CMEs from the upper corona into the heliosphere. SoloHI provides an additional POV that complements the remote sensing observations made by other {heliopheric} missions and incorporates observations at different heliocentric distances closer to the Sun. With a higher resolution and higher cadence than those found in previous generations of HIs, SoloHI allows us to have a better understanding of the CME internal structure. SoloHI also contributes to monitoring different spacecraft and planets, providing new information to the space weather community. 

In this work, we have developed a multi-viewpoint catalog based on the SoloHI events, {but complementing the description of each observed event with the information provided by other heliospheric missions.} This catalog will be updated as the number of observations made by SoloHI increases.

In this first publication, we reported 140 events detected from January 2022 until April 2023. Whenever possible, we tracked the events throughout their evolution from the source region on the Sun until 1$\,$au. Table \ref{tab:sum} shows a summary of the entire dataset available in the catalog.

\begin{table}[]
    \begin{tabular}{c|ccc}
         Quality classification & Good: 73 & Faint: 42 & {Poor}: 25\\
          \hline
        Other RS Observations & LASCO-C2/3: 136 & COR-1/2: 131 & WISPR: 11\\
          \hline
        Source Region & Detected: 88 & Flares: 41 & Filament: 46\\
          \hline
         in-situ detections & Detected: 26 & & \\ 
    \end{tabular}
    \caption{Summary of the main characteristics of the 140 events detected by SoloHI and available in the catalog.}
    \label{tab:sum}
\end{table}


We found 73 events in which we could track different features along their evolution in the entire SoloHI FOV. The rest of the events were classified as faint (42) or {poor} (25) events. We identified 137 events that had been detected by at least one other remote sensing instrument during this period. We identified 88 source regions, of which 41 are associated with flares, and 46 present filaments related to the CMEs. Only 26 events were detected in-situ by at least one spacecraft.

During the first science perihelion, SoloHI served as a complementary viewpoint to LASCO and STA in 16 events, adding important additional information about CME structures from another perspective. SoloHI also observed 6 geo-effective events of which SolO has in-situ observations for two of these. During the second science perihelion, SoloHI was behind the limb from the Earth's perspective and detected 11 events that no other instruments observed. These observed CMEs can also be studied by the complete set of RS instruments available in the SolO mission. 

The goal of this multi-viewpoint catalog is to facilitate scientific research by providing a global description of the events to be analyzed. By linking the different catalogs available online, we join efforts and contribute to different research areas. This catalog will be open to the entire scientific community and will be updated throughout the course of the Solar Orbiter mission.

As a next step, with a larger sample of events, we will perform a statistical analysis on some of the parameters in the catalog and learn how SoloHI is improving the observations and the understanding of CME 3D structure. 

{The current set of observations of the corona and heliosphere provides a wealth of data from multiple spacecraft and missions. These data span multiple viewpoints from the photosphere, corona, heliospheric, and interplanetary space. Many of these missions and observations provide cataloged data of CME events that have been used for decades. The integration of SoloHI data with the greater community of scientific observations and catalogs enables groundbreaking science and facilitates global analysis. The comprehensive SoloHI CME catalog empowers heliospheric research, offering the community a tool for a comprehensive, multi-viewpoint understanding of CMEs.}

\textit{\textbf{Funding Declaration:} Solar Orbiter is a mission of international cooperation between the European Space Agency (ESA) and the National Aeronautics and Space Administration (NASA), operated by ESA. The Solar Orbiter Heliospheric Imager (SoloHI) instrument was designed, built, and is now operated by the US Naval Research Laboratory with the support of the NASA Heliophysics Division, Solar Orbiter Collaboration Office under DPR NNG09EK11I. P.H. and R.C. are also supported by the Office of Naval Research.}

\appendix
\section{Additional Resources}
\label{Appendix}


Table \ref{table:EventsRSW} is a subsection of the events detected during the first two science perihelia in 2022. 

Table \ref{ListSoloHI} gives the start date for all the 140 events observed by SoloHI from November 2021 to April 2023. The complete description of the events is available on the catalog webpage.


\begin{landscape}
\begin{table}
    \caption{List of SoloHI events observed in the first two science perihelia. The table shortlists the details compiled in the catalog, where the hyperlinks and weblinks of the corresponding movies of the events are available on the webpage of the catalog.}
\begin{tabular}{ccc|cccccc|c|cc}
Start       &      &  & Source  &  &  &  &  &  & Model & Other Spacecraft      &   \\
\hline
\\
Date       & Hour     & Type & \#AR  & LocAR         & EUV & Fl & Fi & HMI & DONKI & Remote Sensing       & In-situ           \\
\hline
\\
03 Mar 2022 & 05:49:34 & L    & 12958 & N18E27        & Y                & Y & Y & Y & Y         & LASCO/COR & -                 \\
09 Mar 2022 & 17:07:45 & L    & -     & -             & -                & - & - & - & -         & LASCO/COR & -                 \\
10 Mar 2022 & 07:31:45 & F    & -     & S40E50        & Y                & - & Y & Y & Y         & LASCO/COR & WIND              \\
10 Mar 2022 & 23:31:45 & F    & 12962 & N12W12        & Y                & - & - & Y & Y         & LASCO/COR & WIND - STA - SolO \\
17 Mar 2022 & 00:42:19 & F    & 12967 & N30E20/N38E75 & Y                & - & Y & Y & Y         & LASCO/COR & WIND - STA        \\
19 Mar 2022 & 19:30:19 & O    & -     & S30E25        & Y                & - & Y & Y & Y         & LASCO          & STA          \\
20 Mar 2022 & 13:06:19 & F    & 12971 & N17E11        & Y                & - & Y & Y & Y         & LASCO/COR & -                 \\
22 Mar 2022 & 20:42:19 & F    & -     & -             & -                & - & - & - & -         &  & -   \\
25 Mar 2022 & 07:06:19 & F    & 12974 & S19E36        & Y                & Y & - & Y & Y         & LASCO/COR & WIND - STA        \\
28 Mar 2022 & 13:06:19 & L    & 12975 & N12W01        & Y                & Y & Y & Y & Y         & LASCO/COR & WIND              \\
28 Mar 2022 & 21:30:19 & F    & 12975 & N12W06        & Y                & Y & Y & Y & Y         & LASCO/COR2 & WIND             \\
30 Mar 2022 & 18:58:19 & F    & 12975 & N13W34        & Y                & Y & - & Y & Y         & LASCO/COR & -                 \\
31 Mar 2022 & 22:10:19 & L    & 12975 & N13W48        & Y                & Y & - & Y & Y         & LASCO/COR & -                 \\
02 Apr 2022 & 14:34:20 & F    & 12976 & N15W61        & Y                & Y & Y & Y & Y         & LASCO/COR & SolO              \\
03 Apr 2022 & 17:46:20 & F    & 12981 & S25W28        & Y                & - & Y & Y & Y         & LASCO/COR & -                 \\
04 Apr 2022 & 17:31:00 & L    & 12979 & S20W110       & Y                & - & - & Y & Y         & LASCO/COR & SolO              \\
08 Oct 2022 & 06:46:07 & L    & -     & -             & -                & - & - & - & Y         & LASCO/COR & -                 \\
09 Oct 2022 & 19:10:07 & O    & -     & -             & -                & - & - & - & -         & LASCO/COR & -                 \\
10 Oct 2022 & 18:22:07 & O    & -     & -             & -                & - & - & - & -         & LASCO/COR & -                 \\
14 Oct 2022 & 15:34:07 & F    & -     & -             & -                & - & - & - & -         & -                    & -      \\
20 Oct 2022 & 15:10:20 & L    & -     & -             & -                & - & - & - & -         & -                    & -      \\
20 Oct 2022 & 21:58:20 & O    & -     & S40E105             & Y                & - & - & - & -         & LASCO/COR & -                 \\
23 Oct 2022 & 11:37:44 & L    & -     & -             & -                & - & - & - & Y         & LASCO/COR & -                 \\
26 Oct 2022 & 05:13:44 & O    & -     & S15E90        & Y                & - & - & - & Y         & LASCO/COR & -                 \\
27 Oct 2022 & 03:14:10 & O    & -     & -             & -                & - & - & - & -         & LASCO/COR & -                 \\
28 Oct 2022 & 08:50:10 & L    & -     & -             & -                & - & - & - & -         & LASCO/COR & -                 \\
31 Oct 2022 & 16:50:41 & F    & -     & -             & -                & - & - & - & Y         & LASCO/COR & -                              
\end{tabular}
\label{table:EventsRSW}

\end{table}
\end{landscape}

\begin{landscape}
\begin{table}
    \caption{SoloHI event list from January 2022 up to April 2023. We report here the time of the first observations of the events in the SoloHI FOV. All the detailed information for these events can be found on the catalog webpage.}
\begin{tabular}{|ccc|ccc|ccc|ccc|}
   & DD-MM-YYYY  & Hour    &    & DD-MM-YYYY & Hour    &     & DD-MM-YYYY & Hour    &     & DD-MM-YYYY & Hour     \\
1  & 12 Jan 2022 & 05:45:14 & 36 & 15 Apr 2022 & 03:33:41 & 71  & 18 Aug 2022 & 15:21:07 & 106 & 08 Dec 2022 & 22:09:07 \\
2  & 14 Jan 2022 & 10:09:14 & 37 & 16 Apr 2022 & 15:27:41 & 72  & 19 Aug 2022 & 07:21:07 & 107 & 10 Dec 2022 & 01:21:07 \\
3  & 16 Jan 2022 & 19:45;14 & 38 & 17 Apr 2022 & 11:10:13 & 73  & 20 Aug 2022 & 10:33:07 & 108 & 15 Dec 2022 & 02:57:14 \\
4  & 21 Jan 2022 & 12:09:14 & 39 & 19 Apr 2022 & 08:46:14 & 74  & 26 Aug 2022 & 18:56:41 & 109 & 15 Dec 2022 & 17:21:14 \\
5  & 27 Jan 2022 & 05:21:07 & 40 & 20 Apr 2022 & 05:34:14 & 75  & 27 Aug 2022 & 04:32:41 & 110 & 16 Dec 2022 & 07:45:14 \\
6  & 02 Feb 2022 & 00:57:07 & 41 & 29 Apr 2022 & 07:30:25 & 76  & 28 Aug 2022 & 14:08:41 & 111 & 21 Dec 2022 & 23:45:14 \\
7  & 03 Feb 2022 & 20:09:07 & 42 & 30 Apr 2022 & 18:43:09 & 77  & 29 Aug 2022 & 21:07:52 & 112 & 03 Jan 2023 & 09:21:14 \\
8  & 04 Feb 2022 & 13:44:41 & 43 & 01 May 2022 & 12:19:09 & 78  & 30 Aug 2022 & 21:07:52 & 113 & 09 Jan 2023 & 02:57:14 \\
9  & 08 Feb 2022 & 15:20:41 & 44 & 03 May 2022 & 21:31:09 & 79  & 05 Sep 2022 & 20:19:09 & 114 & 12 Jan 2023 & 01:21:14 \\
10 & 13 Feb 2022 & 02:19:52 & 45 & 11 May 2022 & 01:20:41 & 80  & 08 Sep 2022 & 23:54:25 & 115 & 15 Jan 2023 & 10:57:14 \\
11 & 20 Feb 2022 & 03:55:09 & 46 & 11 May 2022 & 22:08:41 & 81  & 15 Sep 2022 & 13:53:48 & 116 & 26 Jan 2023 & 06:09:14 \\
12 & 20 Feb 2022 & 07:07:09 & 47 & 13 May 2022 & 14:08:41 & 82  & 16 Sep 2022 & 04:17:48 & 117 & 30 Jan 2023 & 14:09:14 \\
13 & 21 Feb 2022 & 11:54:25 & 48 & 18 May 2022 & 00:57:07 & 83  & 20 Sep 2022 & 23:28:41 & 118 & 03 Feb 2023 & 02:57:14 \\
14 & 28 Feb 2022 & 13:53:48 & 49 & 19 May 2022 & 02:33:07 & 84  & 22 Sep 2022 & 17:04:41 & 119 & 05 Feb 2023 & 10:57:14 \\
15 & 03 Mar 2022 & 05:49:34 & 50 & 25 May 2022 & 23:21:14 & 85  & 23 Sep 2022 & 15:52:41 & 120 & 09 Feb 2023 & 17:21:07 \\
16 & 09 Mar 2022 & 17:07:45 & 51 & 05 Jun 2022 & 10:33:14 & 86  & 24 Sep 2022 & 09:28:41 & 121 & 10 Feb 2023 & 23:45:07 \\
17 & 10 Mar 2022 & 07:31:45 & 52 & 09 Jun 2022 & 02:33:14 & 87  & 28 Sep 2022 & 08:15:44 & 122 & 11 Feb 2023 & 15:45:07 \\
18 & 10 Mar 2022 & 23:31:45 & 53 & 16 Jun 2022 & 16:33:14 & 88  & 02 Oct 2022 & 09:51:21 & 123 & 13 Feb 2023 & 15:45:07 \\
19 & 17 Mar 2022 & 00:42:19 & 54 & 24 Jun 2022 & 07:21:14 & 89  & 03 Oct 2022 & 11:27:21 & 124 & 14 Feb 2023 & 15:45:07 \\
20 & 19 Mar 2022 & 19:30:19 & 55 & 25 Jun 2022 & 02:33:14 & 90  & 03 Oct 2022 & 22:39:21 & 125 & 17 Feb 2023 & 22:08:41 \\
21 & 20 Mar 2022 & 13:06:19 & 56 & 26 Jun 2022 & 07:21:14 & 91  & 04 Oct 2022 & 14:39:21 & 126 & 21 Feb 2023 & 06:08:41 \\
22 & 22 Mar 2022 & 20:42:19 & 57 & 29 Jun 2022 & 19:47:14 & 92  & 08 Oct 2022 & 06:46:07 & 127 & 22 Feb 2023 & 17:20:41 \\
23 & 24 Mar 2022 & 14:42:19 & 58 & 03 Jul 2022 & 10:33:14 & 93  & 09 Oct 2022 & 19:10:07 & 128 & 23 Feb 2023 & 02:56:41 \\
24 & 25 Mar 2022 & 07:06:19 & 59 & 05 Jul 2022 & 12:09:14 & 94  & 10 Oct 2022 & 18:22:07 & 129 & 24 Feb 2023 & 17:07:52 \\
25 & 28 Mar 2022 & 13:06:19 & 60 & 09 Jul 2022 & 18:33:14 & 95  & 14 Oct 2022 & 15:34:07 & 130 & 27 Feb 2023 & 10:43:52 \\
26 & 28 Mar 2022 & 21:30:19 & 61 & 17 Jul 2022 & 08:33:14 & 96  & 20 Oct 2022 & 15:10:20 & 131 & 12 Mar 2023 & 07:29:48 \\
27 & 30 Mar 2022 & 18:58:19 & 62 & 19 Jul 2022 & 07:21:14 & 97  & 20 Oct 2022 & 21:58:20 & 132 & 13 Mar 2023 & 04:20:38 \\
28 & 31 Mar 2022 & 22:10:19 & 63 & 24 Jul 2022 & 00:57:14 & 98  & 23 Oct 2022 & 11:37:44 & 133 & 14 Mar 2023 & 02:44:38 \\
29 & 02 Apr 2022 & 14:34:20 & 64 & 29 Jul 2022 & 20:09:14 & 99  & 26 Oct 2022 & 05:13:44 & 134 & 17 Mar 2023 & 17:08:04 \\
30 & 03 Apr 2022 & 17:46:20 & 65 & 31 Jul 2022 & 08:57:14 & 100 & 27 Oct 2022 & 03:14:10 & 135 & 18 Mar 2023 & 10:44:04 \\
31 & 04 Apr 2022 & 17:31:00 & 66 & 14 Aug 2022 & 18;57:14 & 101 & 28 Oct 2022 & 08:50:10 & 136 & 19 Mar 2023 & 01:08:04 \\
32 & 09 Apr 2022 & 15:33:10 & 67 & 15 Aug 2022 & 10:33:14 & 102 & 31 Oct 2022 & 16:50:41 & 137 & 20 Mar 2023 & 16:48:32 \\
33 & 10 Apr 2022 & 07:33:10 & 68 & 16 Aug 2022 & 08:57:07 & 103 & 28 Nov 2022 & 04:49:41 & 138 & 29 Mar 2023 & 18:47:37 \\
34 & 11 Apr 2022 & 07:57:10 & 69 & 16 Aug 2022 & 21:45:07 & 104 & 29 Nov 2022 & 05:37:41 & 139 & 01 Apr 2023 & 09:55:44 \\
35 & 13 Apr 2022 & 10:45:41 & 70 & 17 Aug 2022 & 18:33:07 & 105 & 01 Dec 2022 & 04:32:41 & 140 & 03 Apr 2023 & 00:19:45
\end{tabular}
\label{ListSoloHI}
\end{table}
\end{landscape}

\begin{acknowledgements}
{The EUI instrument was built by CSL, IAS, MPS, MSSL/UCL, PMOD/WRC, ROB, LCF/IO with funding from the Belgian Federal Science Policy Office (BELSPO/PRODEX PEA 4000112292 and 4000134088); the Centre National d’Etudes Spatiales (CNES); the UK Space Agency (UKSA); the Bundesministerium für Wirtschaft und Energie (BMWi) through the Deutsches Zentrum für Luft- und Raumfahrt (DLR); and the Swiss Space Office (SSO).} We acknowledge the Community Coordinated Modeling Center (CCMC) at Goddard Space Flight Center for the use of the CME records provided by the DONKI catalog. The COR1 CME catalog is generated and maintained by the GSFC STEREO COR1 team. The STEREO SECCHI data are produced by an international consortium of the NRL, LMSAL, NASA GSFC (USA), RAL and the University of Birmingham (UK), MPS (Germany), CSL (Belgium), and IOTA and IAS (France). LASCO CME catalog is generated and maintained at the CDAW Data Center by NASA and The Catholic University of America in cooperation with the Naval Research Laboratory. SOHO is a project of international cooperation between ESA and NASA.

\end{acknowledgements}

\end{document}